\documentclass[twocolumn]{aastex631}

\usepackage{times}
\usepackage{amsmath,amssymb,amstext}
\hypersetup{linkcolor=red,citecolor=blue,filecolor=cyan,urlcolor=blue}
\usepackage{tabularx}
\usepackage{longtable}
\usepackage{float}
\usepackage{natbib}
\usepackage{graphicx,color}
\usepackage{txfonts}

\usepackage{multirow}
\usepackage{array}
\usepackage{rotating}
\usepackage{sidecap}
\usepackage{hyperref}
\usepackage{epstopdf}
\usepackage{footnote}
\usepackage{tabularx}
\usepackage{booktabs}
\usepackage{longtable}
\usepackage{tabu}
\usepackage{longtable}
\newcommand{\kms}{km\,s$^{-1}$}

\begin{document}

\title{{\large{\bf Imaging and Radio Signatures of Shock–Plasmoid Interaction}}}
\author{Pankaj Kumar\altaffiliation{1,2}}

\affiliation{Department of Physics, American University, Washington, DC 20016, USA}
\affiliation{Heliophysics Science Division, NASA Goddard Space Flight Center, Greenbelt, MD, 20771, USA}

\author{Judith T.\ Karpen}
\affiliation{Heliophysics Science Division, NASA Goddard Space Flight Center, Greenbelt, MD, 20771, USA}

\author{P. K.\ Manoharan}
\affiliation{Heliophysics Science Division, NASA Goddard Space Flight Center, Greenbelt, MD, 20771, USA}
\affiliation{The Catholic University of America, 20664, Washington, DC, USA}

\author{N. \ Gopalswamy}
\affiliation{Heliophysics Science Division, NASA Goddard Space Flight Center, Greenbelt, MD, 20771, USA}

\email{pankaj.kumar@nasa.gov}

\begin{abstract}
Understanding how shocks interact with coronal structures is crucial for understanding the mechanisms of particle acceleration in the solar corona and inner heliosphere. Using simultaneous radio and white-light observations, we investigate the interaction between a CME-driven shock and a plasmoid. LASCO and STEREO-A COR-2 white-light images are analyzed to track the evolution of the plasmoid, CME and its associated shock, while the Wind/WAVES and STEREO/WAVES dynamic spectra provide complementary radio signatures of the shock-plasmoid interaction at $\approx$7 R$_\odot$. An interplanetary Type II radio burst was detected as the shock propagated through the plasmoid. The merging of the plasmoid into the CME was accompanied by interplanetary Type III radio bursts, suggesting escaping electron beams during the reconnection process. These observations clearly demonstrate that shock-plasmoid interactions can enhance the efficiency of particle acceleration associated with CMEs, with implications for electron acceleration  in flare and heliospheric current sheets as well.
\end{abstract}

\section{INTRODUCTION}\label{intro}
Coronal mass ejections (CMEs) are among the most energetic transient phenomena in the heliosphere, capable of driving shocks, accelerating particles, and triggering geomagnetic storms. When multiple CMEs are launched in close succession with different speeds, they can interact in interplanetary space. Such CME--CME interactions may lead to shock merging, magnetic field compression, and complex particle acceleration processes, making them critical for understanding space weather impacts \citep{Gopalswamy2001, Gopalswamy2002, Lugaz2005}. Observational signatures of CME--CME interactions appear in both white-light and radio observations. In coronagraph images, they manifest as deflections, merging structures, or sudden kinematic changes \citep{Lugaz2012, Vourlidas2013}. Radio observations often reveal complex features in dynamic spectra. Type II bursts, associated with CME-driven shocks, can exhibit enhanced emission during CME-CME interaction \citep{Gopalswamy2001,Gopalswamy2002}. Such interactions can enhance intensity of shock-driven particle acceleration, producing intense SEP events \citep{Li2012, Kahler2014}.

Magnetic reconnection in solar flares occurs within a current sheet beneath the erupting flux rope, where dynamic processes such as plasmoid formation and shock development can take place \citep{Shibata1995, Lin2005,karpen2012}. Plasmoids, formed via tearing-mode instability, are key signatures of fast and bursty reconnection, and they facilitate efficient energy release and particle acceleration \citep{shibata2001, loureiro2007, Bhattacharjee2009}. Termination shocks can develop when reconnection outflows collide with closed magnetic loops constituting the flare arcade \citep{Forbes1986, Tsuneta1997, Chen2015}. These shocks are believed to play a crucial role in accelerating high-energy particles, particularly electrons observed in hard X-rays and radio emissions \citep{Mann2009}. 
Despite their potential significance, interactions between plasmoids and shocks remain underexplored in flare current sheets. Shock-plasmoid interactions and coalescence of plasmoids can enhance particle acceleration \citep{Drake2006, Guo2015, Kumar2025}. A detailed understanding of shock–plasmoid interactions is therefore essential for understanding particle acceleration in solar flares as well as in CMEs.

In this letter, we report a clear example of an interaction between a CME-driven shock and a small blob (interpreted as a plasmoid), observed in LASCO C2 and STEREO-A COR2 coronagraph images. To our knowledge, such a shock–plasmoid interaction has not been reported previously.
We observed distinct radio signatures associated with the interaction between a shock and a slowly moving blob ahead of it. In the dynamic radio spectra from WIND/WAVES and STEREO/WAVES, we identified a Type II radio burst that occurred at the same time as the shock transited the blob in the coronagraph images. The observed radio and white-light signatures provide clear evidence of CME-driven shock and plasmoid interaction, accompanied by enhanced electron acceleration. Moreover, interplanetary Type III radio bursts were observed during the merger between the blob and the CME, indicating the formation of electron beams during this interaction. Our analysis yields estimates of the ambient magnetic-field strength and electron density in the blob, identifies the probable source of the blob, and suggestions for broader applications of shock-plasmoid encounters.  

\section{Data}\label{data}
We analyzed {\it Solar Dynamics Observatory} (SDO)/Atmospheric Image Assembly (AIA; \citealt{lemen2012}) full-disk images of the Sun (field-of-view $\approx$1.3~R$_\odot$) with a spatial resolution of 1.5$\arcsec$ (0.6$\arcsec$~pixel$^{-1}$) and a cadence of 12~s, in the following channels: 211~\AA\ (\ion{Fe}{14}, $T\approx 2$~MK), 131~\AA\ (\ion{Fe}{8}, \ion{Fe}{21}, \ion{Fe}{23}, i.e., 0.4, 10, 16 MK) images.
 A 3D noise-gating technique \citep{deforest2017} was used to clean the SDO AIA images.

The Extreme Ultraviolet Imager \citep[EUVI;][]{Wuelser2004,Howard2008} on STEREO {\rm Solar TErrestrial RElations Observatory} observed the studied flare behind the east limb. The separation angle between SDO and STEREO-A (STEREO-B) was 154.6$^{\circ}$(-163.5$^{\circ}$) on March 29 2014. 
We used STEREO-A COR1 (1.3-4 $R_{\sun}$) and COR2 (2.0-15 $R_{\sun}$) \citep{thompson2003} and SOHO (Solar and Heliospheric Observatory) LASCO C2 (2-6 $R_{\sun}$) \citep{brueckner1995,yashiro2004} coronagraph images for CME propagation in the interplanetary medium. 

We utilized dynamic radio spectra obtained by the Radio Solar Telescope Network (RSTN) Learmonth Radio Observatory for metric/decimetric emission from the low corona, and WIND/WAVES \citep{bougeret1995} for the interplanetary medium. We used Wind 3DP (Three-Dimensional Plasma and Energetic Particles: \citealt{lin1999}) and SOHO ERNE (Energetic and Relativistic Nuclei and Electron: \citealt{torsti1995}) observations for the SEPs (electrons/protons) at 1 AU.

\begin{figure*}
\centering{
\includegraphics[width=18cm]{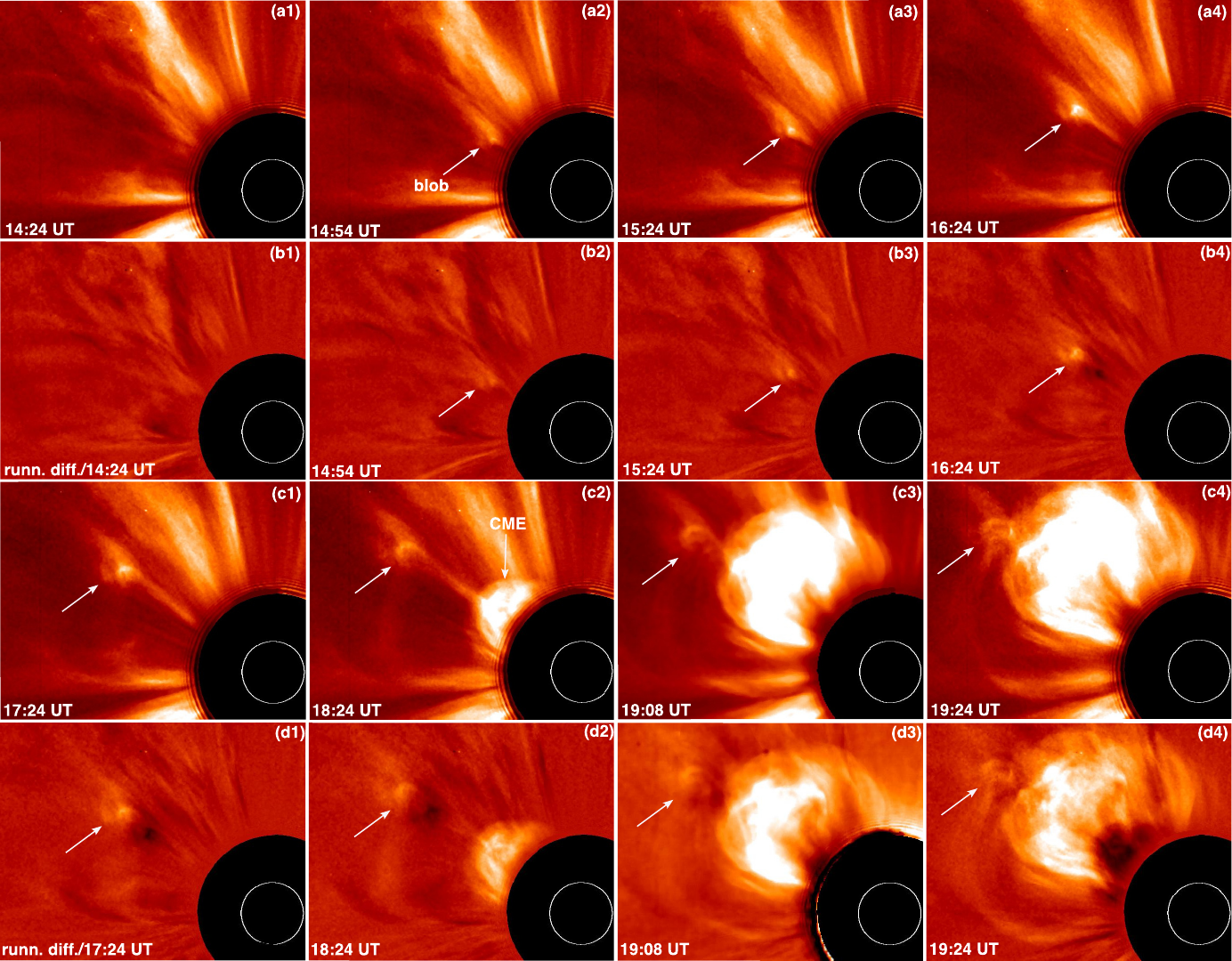}
}
\caption{CME-blob interaction. STEREO-A COR-2 coronagraph intensity (rows 1 and 3) and running-difference (rows 2 and 4) images showing a slowly moving blob (marked by arrows) and its interaction with a CME and its shock.} 
\label{fig1}
\end{figure*}
\begin{figure*}
\centering{
\includegraphics[width=18cm]{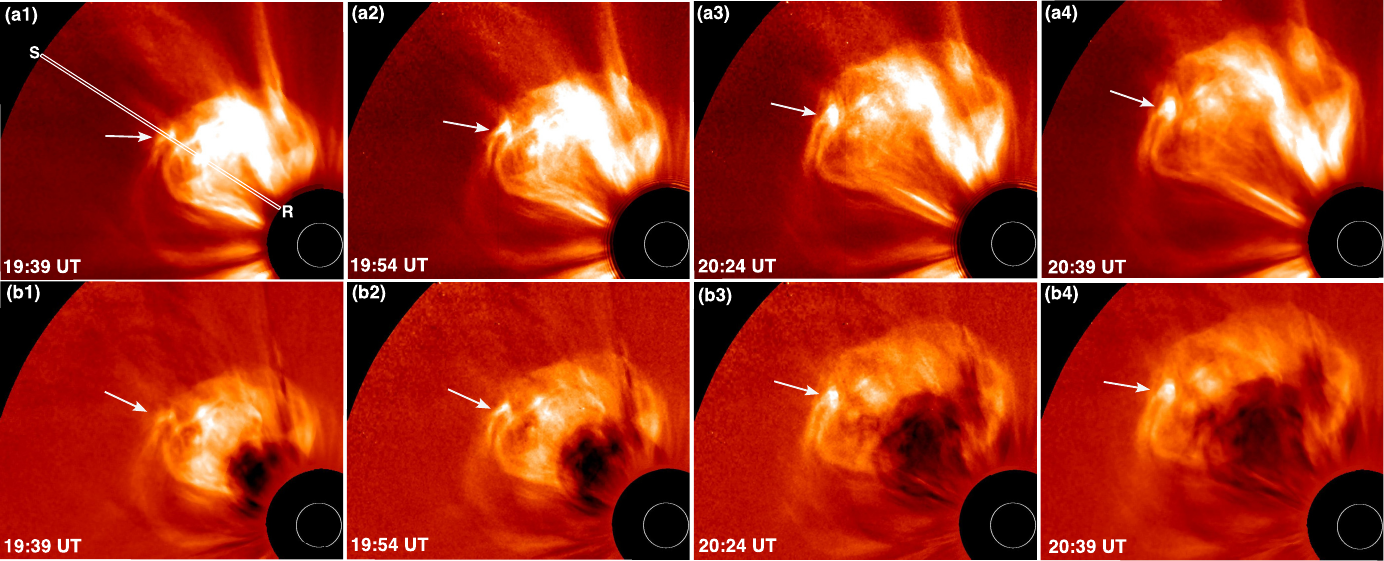}
}
\caption{Coalescence of the blob into the CME. STEREO-A COR2 coronagraph intensity (top row) and running-difference (bottom row) images showing the interaction between the shock and the blob, and subsequent merging of the blob (marked by arrows) into the CME. The slit RS marked in panel (a1) is used to generate the time–distance intensity plot shown in Figure \ref{fig4}.} 
\label{fig2}
\end{figure*}
\begin{figure*}
\centering{
\includegraphics[width=18cm]{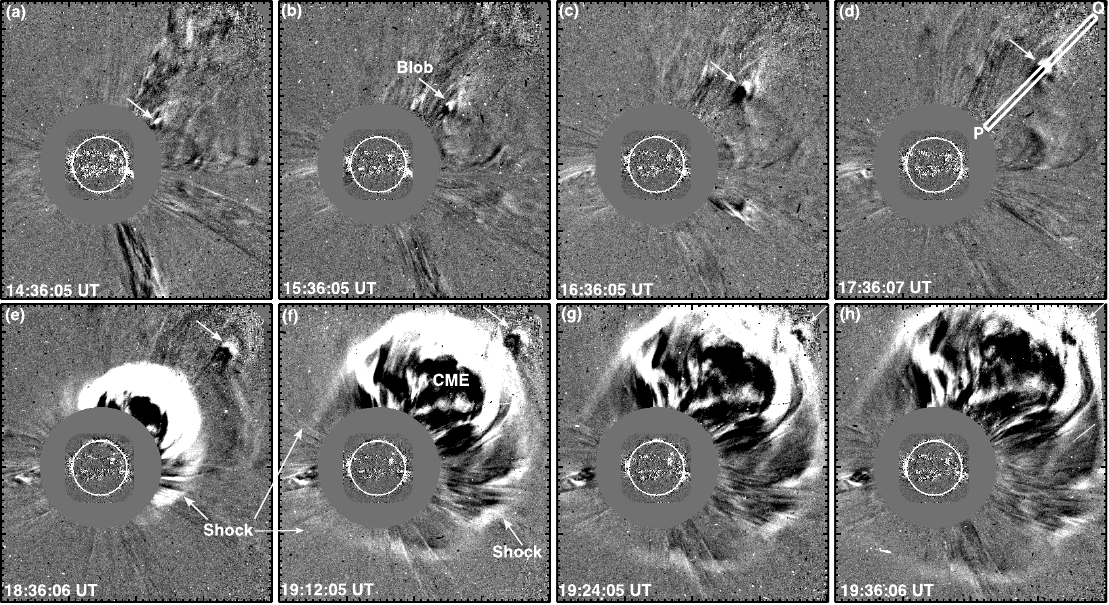}
}
\caption{CME/shock-blob interaction in LASCO C2. A sequence of LASCO C2 coronagraph running-difference images leading up to the interaction at 19:24-19:36 UT between the CME shock and the slowly moving blob (indicated by arrows). The slit PQ marked in panel (d) is used to generate the time–distance intensity plot shown in Figure \ref{fig4}.} 
\label{fig3}
\end{figure*}

\section{Results}\label{res}

An X1.0 flare occurred in AR NOAA 12017 (located at N10W32) on March 29, 2014. The associated eruption took place within a closed fan-spine topology embedded under the western lobe of a pseudostreamer—commonly referred to as a nested null-point topology \citep{karpen2024}—with inner and outer nulls labeled N1 and N2, respectively (Figure \ref{app-fig1} (a)). According to the GOES soft X-ray flux, the flare began at $\approx$17:35~UT, peaked at $\approx$17:48~UT, and ended at $\approx$17:54~UT. A flux rope (FR) erupted along with an S-shaped filament located at the polarity inversion line (PIL) (Figure \ref{app-fig1} (b)). The eruption, associated with breakout reconnection at the inner null N1, generated a circular ribbon around the PIL and a remote ribbon at the footpoint of the outer spine (see \citealt{kumar2025b} for more details). A fast EUV wave appeared ahead of the flux rope (Movie S1), and a simultaneous metric Type II radio burst was observed in the dynamic radio spectrum (Figure \ref{app-fig1} (c,d,f)). The projected speed of the EUV wave was approximately 1000~km~s$^{-1}$, closely matching the shock speed estimated from the drift rate of the Type II burst. We interpret the fast EUV wave as a shock driven by the erupting flux rope. The projected height of the shock above the flare center was about 0.15 R$_\odot$ at 17:47:25 UT. The further extension of this shock was observed in LASCO C2 coronagraph images, and an associated gradual solar energetic particle (SEP) event was detected at 1~AU by Wind/3DP. The CME-shock and blob interaction was observed by SOHO/LASCO-C2 and STEREO-A/COR-2 from different viewing angles (Figure \ref{app-fig1} (e)).

STEREO COR2 images reveal the appearance of a blob at approximately 14:54 UT (Figure \ref{fig1}(a2,b2)). Both intensity and running-difference images show the blob slowly moving and expanding between 15:24 and 17:24 UT (Figure \ref{fig1}(a3,a4,b3,b4,c1,d1)). The CME appears at 18:24 UT, behind the blob, and the interaction between the CME-driven shock (visible as a faint structure) and the blob begins during 19:09–19:24 UT (Figure \ref{fig1}(c2-c4,d2-d4)). Later, we observe the coalescence of the blob into the CME (19:39-20:39 UT), accompanied by white-light intensity enhancement during the merging process (Figure \ref{fig2}).

LASCO C2 coronagraph images illustrate the slowly moving blob and its interaction with the CME and its shock from a different viewing angle. The tiny blob first appeared in the LASCO C2 field of view around 14:36 UT and moved slowly outward (Figure \ref{fig3}(a-d)). The CME emerged along the same trajectory at approximately 18:00 UT. A large-scale shock structure ahead of the CME became evident between 18:36 and 19:12 UT (Figure \ref{fig3}(e,f)). A distinct separation between the CME leading edge (frontal loop) and the shock was observed at 19:24 UT, followed by the interaction between the shock and blob during 19:24–19:36 UT (Figure \ref{fig3}(g,h)).

\begin{figure*}
\centering{
\includegraphics[width=18cm]{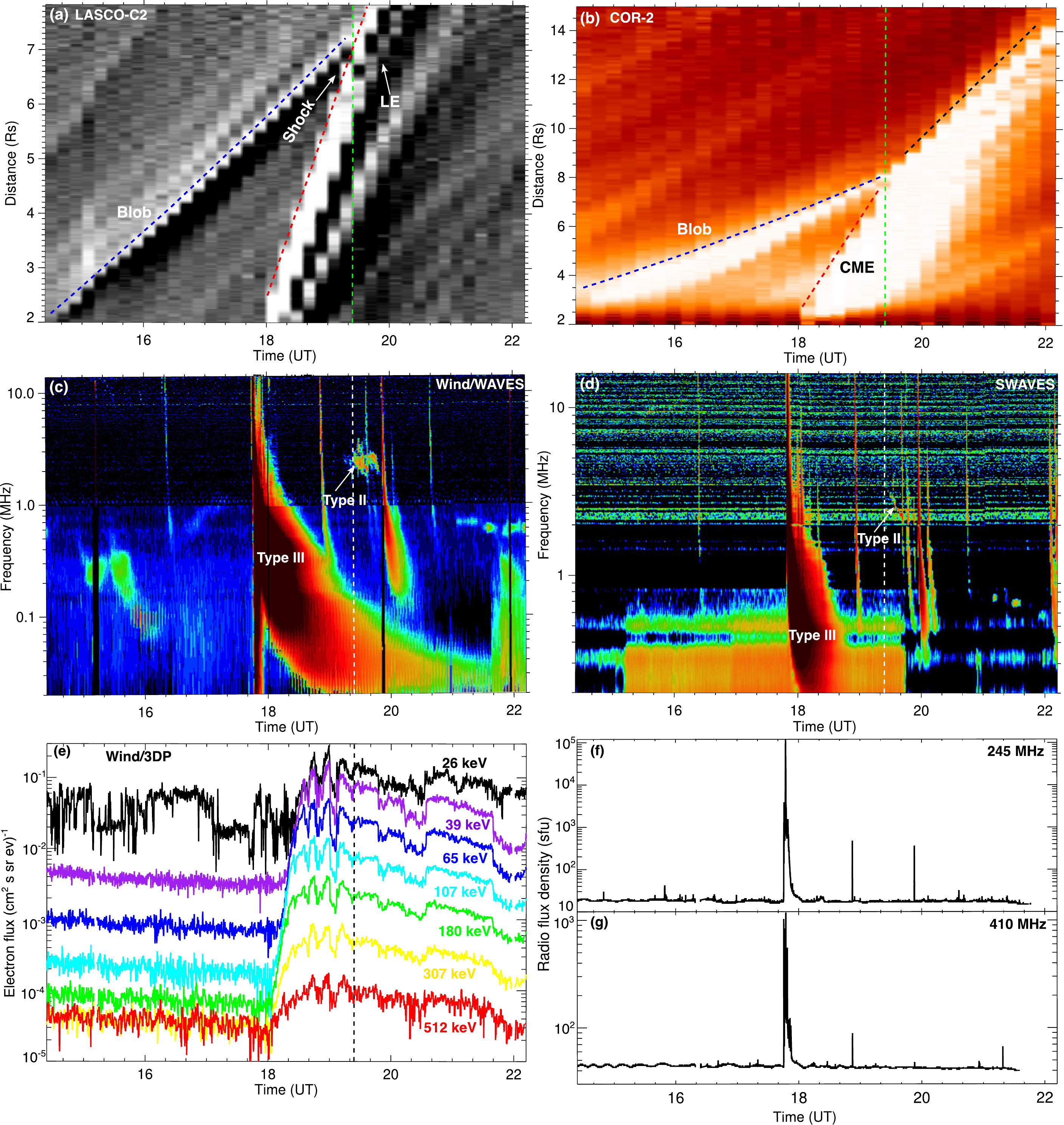}
}
\caption{Kinematics of the blob, CME/shock, and associated radio signatures. (a,b) Time–distance intensity plot along the slits PQ (shown in Figure \ref{fig3}(d)) and RS (shown in Figure \ref{fig2}(a1)) using LASCO C2 and STEREO-A COR-2 images. The vertical green line marks the onset of the interaction between the CME-driven shock and the slowly moving blob at $\approx$19:24 UT. LE indicates the leading edge of the CME. The blue dashed curve represents the second-order polynomial fit to the blob, while the red dashed curve shows the linear fit to the CME shock. (c, d) Dynamic radio spectra from Wind/WAVES (0.02–13.85 MHz) and STEREO-A/WAVES (0.2–16 MHz). The vertical white line marks the onset of the interaction between the CME-driven shock and the slowly moving blob.  (e) SEP electron flux (26–512 keV) observed by Wind/3DP at 1 AU. (f, g) RSTN radio flux density profiles (sfu) at 245 and 410 MHz from the RSTN Sagamore Hill radio observatory. The vertical white line marks the onset of the interaction between the CME-driven shock and the slowly moving blob.} 
\label{fig4}
\end{figure*}

The time–distance intensity plot along the slit P1Q1 using LASCO C2 images (marked in Figure \ref{fig3}(d)) shows the kinematics of the blob and the CME-driven shock overtaking it. A clear separation between the CME leading edge and the shock is visible at 19:24 UT. A second-order polynomial fit was applied to the blob’s height–time measurements (Figure \ref{fig4}(a)), indicating a speed (plane of sky) increase from 187 to 224 \kms and an estimated acceleration of $\approx$2.1 m s$^{-2}$. A linear fit to the shock front yields a speed of about 640 \kms. The interaction between the CME shock and the blob initiates at about 7 R$_\odot$ during 19:24–19:36 UT.

The time–distance intensity plot along slit RS, derived from STEREO-A COR2 intensity images (2–15 R$_\odot$), reveals a slowly moving blob between 14:24 and 19:24 UT, followed by a faster CME around 18:00 UT (Figure \ref{fig4}(b)). A second-order polynomial fit to the blob’s trajectory yields a sky-plane speed ranging from 154 to 215 \kms and an acceleration of $\approx$3.5 m s$^{-2}$.
The CME speed, estimated using a linear fit, is 755 \kms. Due to the different viewing geometry of STEREO-A, a distinct separation between the CME leading edge and the shock clearly visible in LASCO C2 is not clear here. However, the interaction between the CME and the blob is clearly observed at 7 R$_\odot$, consistent with the LASCO C2 measurements. The blob’s white-light intensity increased during the interaction, and its speed after the interaction rose to $\approx$480 \kms.

\begin{figure}
\centering{
\includegraphics[width=9cm]{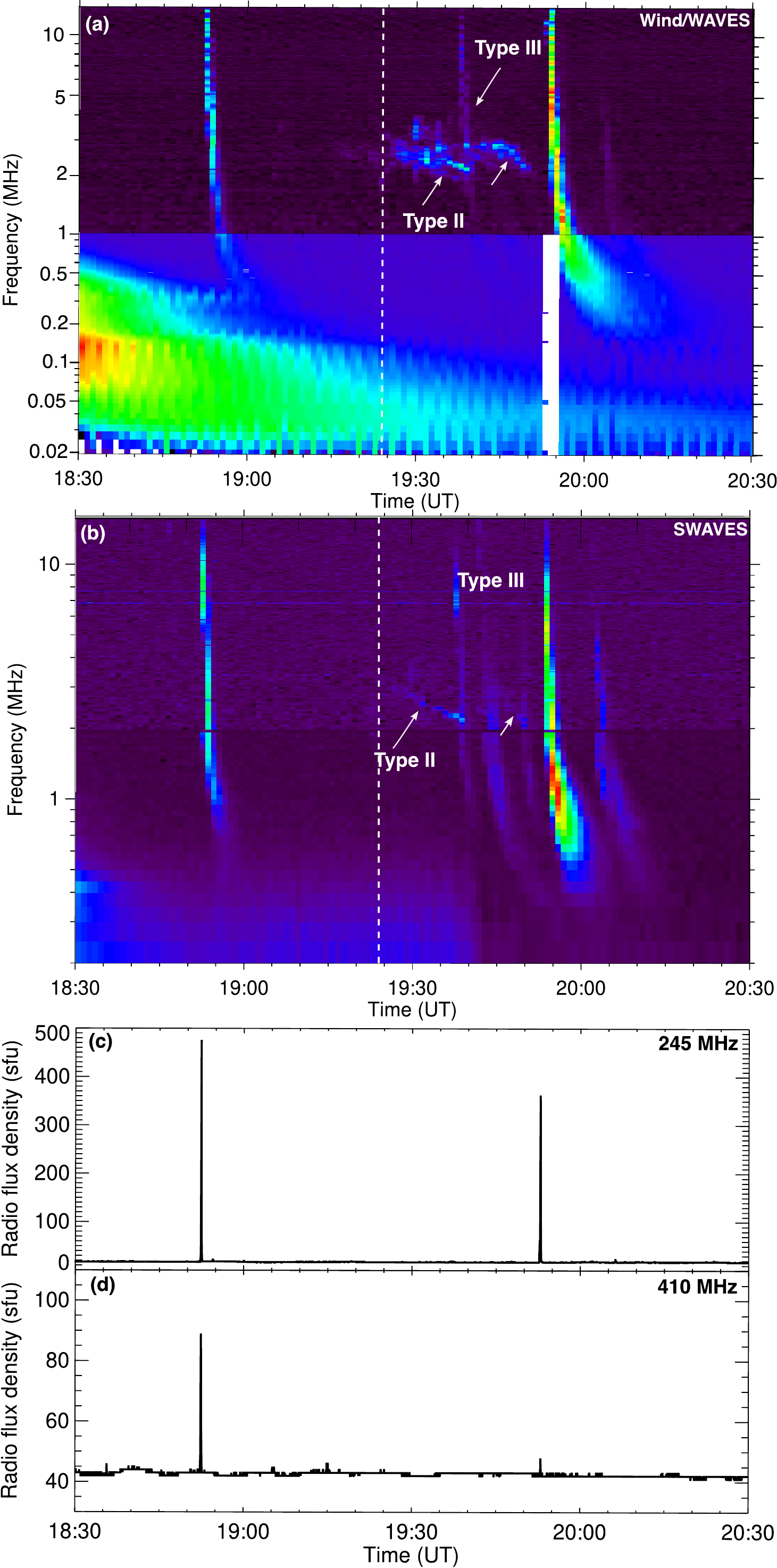}
}
\caption{Radio bursts during the CME/shock-blob interaction. (a,b) Zoomed view of radio bursts detected during the interaction and the merging of the blob into the CME. (c,d) Radio flux density (sfu) profiles at 245/410 MHz (metric/decimetric frequencies) from the RSTN Sagamore Hill radio observatory.} 
\label{fig5}
\end{figure}

The Wind/WAVES and STEREO-A/WAVES dynamic radio spectra reveal a Type II radio burst at 19:24-19:50 UT, during the interaction (Figure \ref{fig4}(c,d)). Figure \ref{fig5} presents a zoomed-in view of the radio bursts observed between 18:30 and 20:30 UT. The band splitting of the Type II burst is clearly visible in the Wind/WAVES dynamic spectrum (Figure \ref{fig5}(a)), with faint splitting features also evident in the STEREO-A/WAVES spectrum (Figure \ref{fig5}(b)). 

The Type II burst was detected in the frequency range of $2$--$3~\mathrm{MHz}$, with a measured frequency drift rate of $-0.00122061~\mathrm{MHz~s^{-1}}$ (Figure \ref{fig5}(b)). To estimate the shock speed, we assumed the plasma emission mechanism where the plasma frequency $f_p$ is related to the local electron density. We calculated the shock speed $V_{\mathrm{shock}}$ using the following relation: $V_{\mathrm{shock}} = \frac{2L}{f_p} \left| \frac{df}{dt} \right|$, where $L$ is the density scale height, $f_p$ is the central plasma frequency, and $df/dt$ is the frequency drift rate from the fundamental band. Since the shock was interacting with a dense blob, we approximate the density scale height by the size of the blob itself. The coronagraph images provide a blob width of about $1~R_\odot$ shortly before the interaction (Figure \ref{fig1}(c3)). Hence the estimated shock speed is $V_{\mathrm{shock}}\approx684~\mathrm{km~s^{-1}}$.
This estimate is consistent with the shock speeds independently measured from coronagraph observations, supporting the interpretation that the Type II burst was generated during the shock-blob interaction.

 Assuming the burst represents plasma emission at the fundamental frequency of 2-3 MHz, the electron density is approximately $7.8\times 10^{4}$ cm$^{-3}$. Here $f_p \approx 9 \sqrt{n_e}$\,kHz, where $f_p$ is the central plasma frequency and $n_e$ is the electron density in cm$^{-3}$. To quantify the physical conditions at the shock–plasmoid interaction near \( \approx 7\, R_\odot \), we analyze the observed band splitting in the fundamental component of the Type II radio burst. The frequency ratio between the upper and lower split bands (\( f_{\text{upper}} = 3.0 \, \text{MHz} \), \( f_{\text{lower}} = 2.5 \, \text{MHz} \)) yields a density compression ratio 
$X = (\frac{f_{\text{upper}}}{f_{\text{lower}}})^2=1.44$.
Assuming a quasi-perpendicular shock geometry, the Alfvén Mach number is estimated using the Rankine–Hugoniot relation \citep{vrsnak2002} $M_A = \sqrt{\frac{X(X + 5)}{2(4 - X)}}\approx 1.35$. The shock speed measured from the coronagraph images is approximately \( v_{\text{shock}} \approx 640 \, \text{km s}^{-1} \), implying an Alfvén speed of $v_A = \frac{v_{\text{shock}}}{M_A} \approx 475 \, \text{km s}^{-1}.$
Using an electron density of \( n_e = 7.8 \times 10^4 \, \text{cm}^{-3} \) within the plasmoid, the estimated magnetic field strength is $B = v_A \sqrt{4\pi m_p n_e}\approx 60~\mathrm{mG}.$ These results suggest a moderately strong shock capable of generating the observed radio emission.

In addition, Wind/3DP detected energetic electrons (26–512 keV) during 18:00–23:00 UT, indicating an ongoing SEP event associated with the CME-driven shock that accompanied the X1.0 flare (Figure \ref{fig4}(e)). Several interplanetary Type III radio bursts were observed during the shock–blob interaction and merger with the CME; however, not all were related to the shock-blob encounter. 
Our analysis of the RSTN radio flux density measurements at metric/decimetric frequencies (245/410 MHz) confirmed the independent solar origin of some of the interplanetary Type III bursts during the interaction phase (Figure \ref{fig4}(f, g)). Notably, strong Type III bursts around 18:53 UT and 19:54 UT were detected in the metric/decimetric ranges (245/410 MHz), indicating their origin in the low corona (Figure \ref{fig5}(c)). 
 On the other hand, the fainter complex Type III bursts observed between 19:24 and 19:50 UT were most likely associated with the interaction and merging of the blob into the CME.
  After the interaction, a strong type III burst observed at 19:54 UT seems to have a low-coronal origin. This type III burst is temporally correlated with a tiny jet from the nearby AR 12014. A weak type III burst around 20:04 UT is likely associated with a jetlet in AR 12014, or it could be related to the blob-merging process. Radio imaging observations are required for further confirmation.

\section{DISCUSSION}\label{discussion}

The interaction between a preceding plasmoid with CME-driven shock and the CME itself in the interplanetary medium, observed in white-light coronagraph images and supported by radio signatures, has important implications for particle acceleration. As the shock encounters the plasmoid, it can compress the magnetic structure, enhancing field gradients and enabling Fermi- or betatron-type particle energization. This also leads to localized plasma heating. Then, as the CME encounters the plasmoid, the CME and plasmoid flux systems can reconnect. This process may increase turbulence downstream of the shock, creating favorable conditions for second-order Fermi acceleration, where trapped particles can gain energy as they scatter between the shock and plasmoid, explaining the observed Type II radio burst. The local Alfv\'en speed is typically low in pre-existing high-density coronal structures such as pseudostreamers and blobs, strengthening the shock and increasing efficiency for electron acceleration as it passes through these high-density regions \citep{cho2013,kumar2016,Kouloumvakos2021,kumar2024}.

Based on AIA 211~\AA\ and STEREO COR1 observations, the blob originated as a result of interchange reconnection near the outer null of the nested fan-spine topology. The associated reconnection signatures included inflows (expansion of the western lobe of the pseudostreamer) toward the outer null, followed by downflows and the development of an outward-moving structure.
The coronagraph images reveal magnetic structures connected to the plasmoid: a bright plasma sheet below it, a structure extending through the plasmoid toward the south, and a faint compression front ahead of the blob.  These observations indicate that the magnetic structure of the plasmoid is a mini flux rope, which is consistent with 3D structure of plasmoids as predicted in MHD simulations of magnetic reconnection \citep[e.g., ][]{daughton2011,dahlin2025}. Therefore, there should be some similarities between the event reported here and CME-CME interactions. 
Unlike CME-CME interactions, however, where both CMEs often drive shocks that merge and create complex radio signatures \citep{Gopalswamy2001,Gopalswamy2002a,luga2017}, the slowly moving plasmoid here lacks a shock, so the observed radio enhancement must result from the CME shock-plasmoid interaction alone. The Type II burst observed during the shock-plasmoid interaction provides direct evidence that shock-driven particle acceleration can be significantly influenced by pre-existing coronal structures. 

Moreover, complex interplanetary Type III radio bursts were observed during the merger between the plasmoid and the CME. These bursts likely result from electrons accelerated at the reconnection site, escaping along newly open magnetic field lines into the heliosphere.
The electron flux detected at 1 AU was slightly enhanced during the plasmoid–CME interaction, suggesting a possible contribution from electron beams propagating into interplanetary space.

\section{CONCLUSIONS}\label{conclusions}

We reported direct imaging of the interaction between a blob (or plasmoid) and a CME-driven shock followed by the CME itself. The shock-plasmoid interaction produced radio signatures of shock-accelerated electrons (Type II radio burst) and escaping electron beams (Type III radio bursts) during the merging of the plasmoid into the CME.
Understanding shock-plasmoid interactions is crucial for improving models of SEP events, as these interactions can enhance particle-acceleration efficiency and may influence SEP properties. Pre-existing plasmoids in the interplanetary medium formed by prior eruptions, or denser coronal structures such as helmet or pseudo-streamers, may serve as important sites for additional particle acceleration beyond the initial CME-driven shock. Future studies using multi-spacecraft observations and simulations will be essential to quantify these effects and refine SEP forecasting models to accommodate these small-scale inhomogeneities. 

Blobs are frequently ejected from streamers into the solar wind \citep{sheeley2009,viall2015, deforest2018}, making their interaction with CMEs and their shocks a common occurrence. We speculate that these interactions should be observed more often in coronagraph images with high temporal/spatial resolution, particularly with the latest missions like Polarimeter to UNify the Corona and Heliosphere (PUNCH), Aditya-L1, CODEX, Solar Orbiter, and Parker Solar Probe. The upcoming SunRISE (Sun Radio Interferometer Space Experiment)) mission will provide the radio imaging of interplanetary radio bursts (100 kHz-20 MHz) produced during the interaction. These missions will significantly improve our ability to study the role of streamer blobs and shock/CME interactions. 

We propose that a similar shock-plasmoid interaction mechanism is likely to operate in flare and heliospheric current sheets. In flare reconnection regions, plasmoids form due to tearing-mode instabilities, while shocks generated by reconnection outflows may interact with these plasmoids, further accelerating particles through interactions with multiple magnetic islands and shocks \citep{shibata2001}. Such interactions might occur in flare current sheets if both plasmoids and shocks are present during magnetic reconnection. MHD simulations of magnetic reconnection in a flare current sheet exhibit small shock fronts ahead of multiple plasmoids \citep{karpen2012}, which may interact with slowly moving plasmoids in the dense plasma sheet. In addition, \citet{nishizuka2013} proposed a model in which plasmoid-shock interactions can enhance particle acceleration efficiency, providing an alternative to conventional shock and reconnection-based acceleration models. Fast-mode waves are frequently observed in EUV images during explosive flare reconnection \citep[e.g., ][]{kumar2017}; they can propagate through the dense current sheet below the erupting flux rope, enhancing the efficiency of electron acceleration and associated modulation of radio emission. This mechanism remains to be tested by future observations along with 3D MHD simulations. Future research integrating white-light, radio, and in-situ observations will be crucial for unraveling the complexities of shock-plasmoid interactions and their role in space weather.\\
\\

\noindent
{\bf {Acknowledgements}}\\
SDO is a mission for NASA's Living With a Star (LWS) program. The LASCO CME catalog is generated and maintained at the CDAW Data Center by NASA and The Catholic University of America in cooperation with the Naval Research Laboratory. SOHO is a project of international cooperation between ESA and NASA. We thank R. Lin and S. Bale (UC Berkeley) and CDAWeb for providing the SEP data. This research was supported by NSF SHINE grant, NASA’s Heliophysics Supporting Research, and Guest Investigator (\#80NSSC20K0265) programs.\\
 \\

\bibliographystyle{aasjournal}
\bibliography{reference.bib}

\clearpage
\appendix
\counterwithin{figure}{section}
\section{Supplementary figures}
\begin{figure*}[htp]
\centering{
\includegraphics[width=15cm]{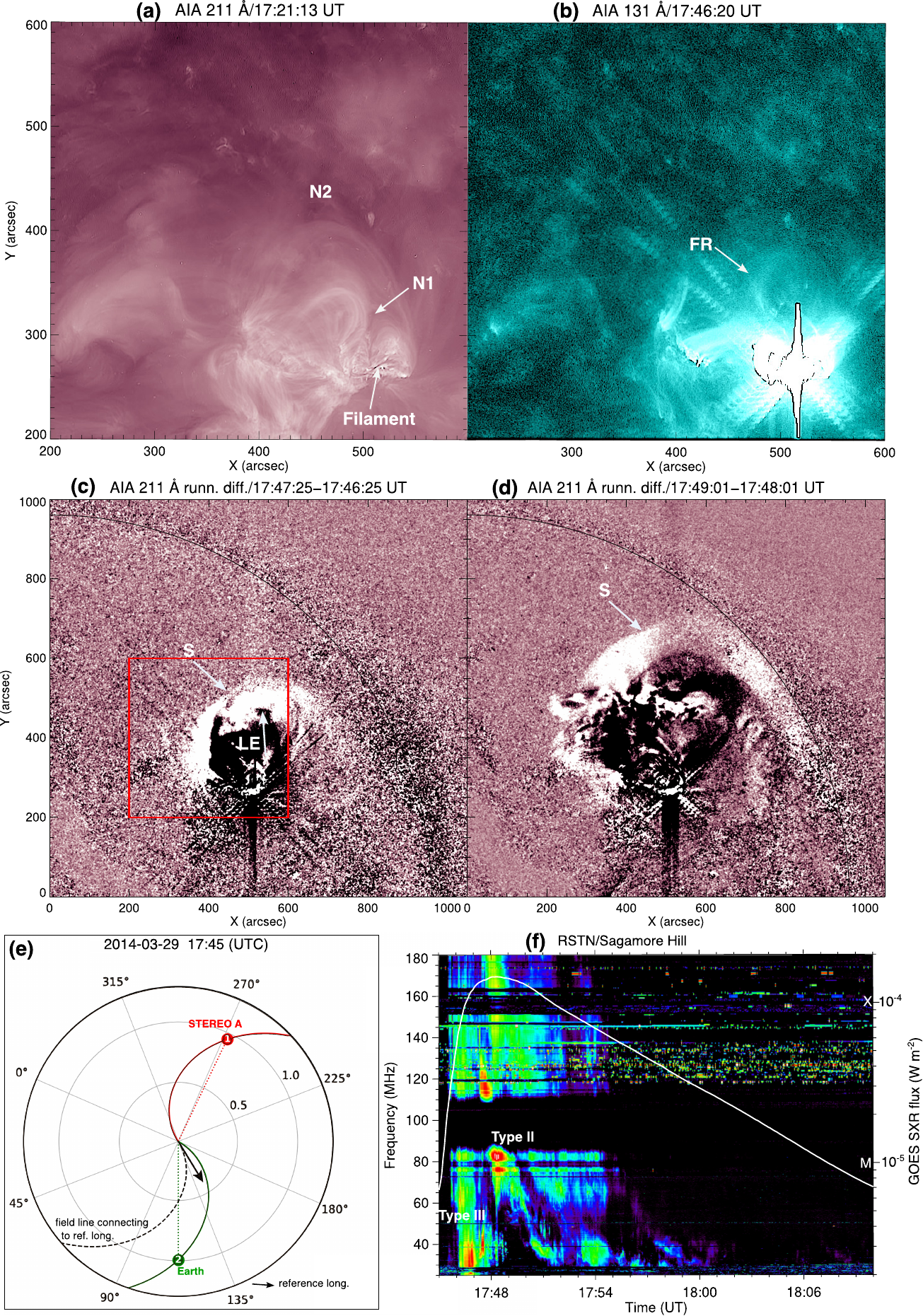}
}
\caption{Eruption and associated fast EUV (shock) wave. 
(a) AIA 211~\AA\ image showing the active region with a nested null-point topology. N1 and N2 mark the approximate positions of the inner and outer nulls, respectively. 
(b) AIA 131~\AA\ image capturing the eruption of the hot flux rope (FR). 
(c,d) AIA 211~\AA\ running-difference images displaying the fast EUV wave (marked by S) ahead of the leading edge (LE) of the flux rope. The red box in (c) indicates the size of the top-row panels.
(e) Magnetic connectivity map for Earth and STEREO-A on March 29, 2014, at 17:50~UT. The arrow indicates the direction of the eruption. 
(f) Dynamic radio spectrum in the 25--180~MHz range from the RSTN Sagamore Hill Radio Observatory. The right Y-axis shows the GOES soft X-ray flux profile in the 1--8~\AA\ channel. } 
\label{app-fig1}
\end{figure*}

This appendix contains additional figures to support the results described above. 
Figure~\ref{app-fig1} shows AIA 211 and 131~\AA\ images of the flux rope eruption from the nested null-point topology and the associated fast EUV (shock) wave. The magnetic connectivity of the spacecraft is determined using the Solar-MACH (Solar Magnetic Connection Haus; \citealt{gieseler2023}) tool.

\begin{figure*}[htp]
\centering{
\includegraphics[width=14cm]{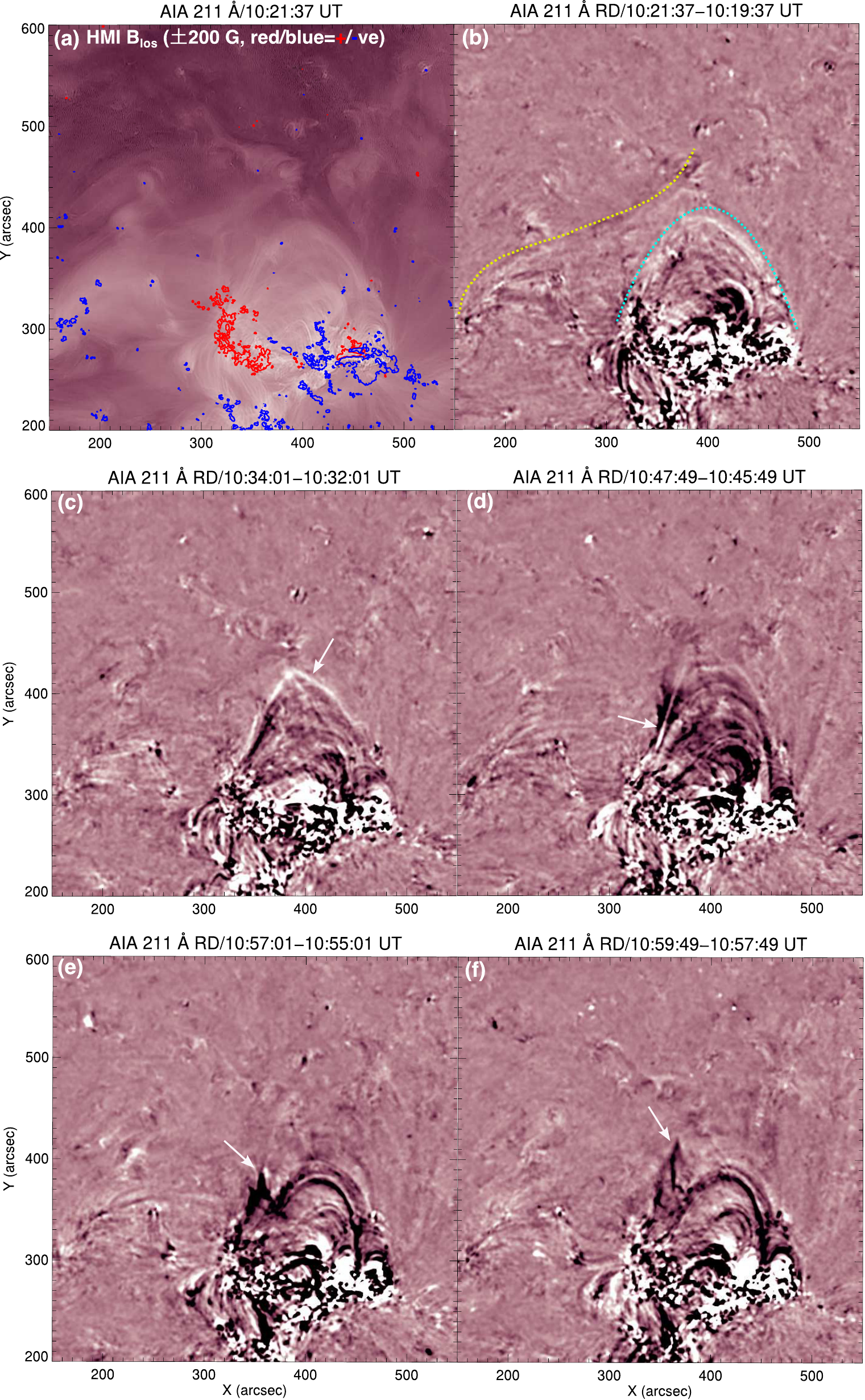}
}
\caption{Blob origin. 
(a) AIA 211~\AA\ image overlaid by HMI magnetogram contours showing the active region with a nested null-point topology. (b-f) AIA 211~\AA\ running-difference images demonstrating the expansion of the loop system on the right side (cyan dashed curve above the nested null N1 and arrow in panel (c)) of the pseudostreamer and associated reconnection between cyan and yellow structures. The arrow in panel (d) indicates the downflow, while the arrows in the bottom panels (e, f) reveal the upward-moving structure connected to the blob observed in STEREO COR-1.} 
\label{app-fig2}
\end{figure*}

The AIA 211~\AA\ images reveal the expansion of the western lobe of the pseudostreamer (cyan dashed curve above the nested fan-spine topology, Figure~\ref{app-fig2}(a,b)) at around 10:20~UT (see Movie~S2). This structure interacted with the feature marked by the yellow dashed curve near the outer null N2. Downflows and an upward-moving structure were observed between 10:45 and 11:00~UT (Figure~\ref{app-fig2}(c-f)). The speed of the upward-moving structure was about 145~km~s$^{-1}$. Therefore, the outward-moving structure was likely produced via interchange reconnection near N2.

\begin{figure*}[htp]
\centering{
\includegraphics[width=14cm]{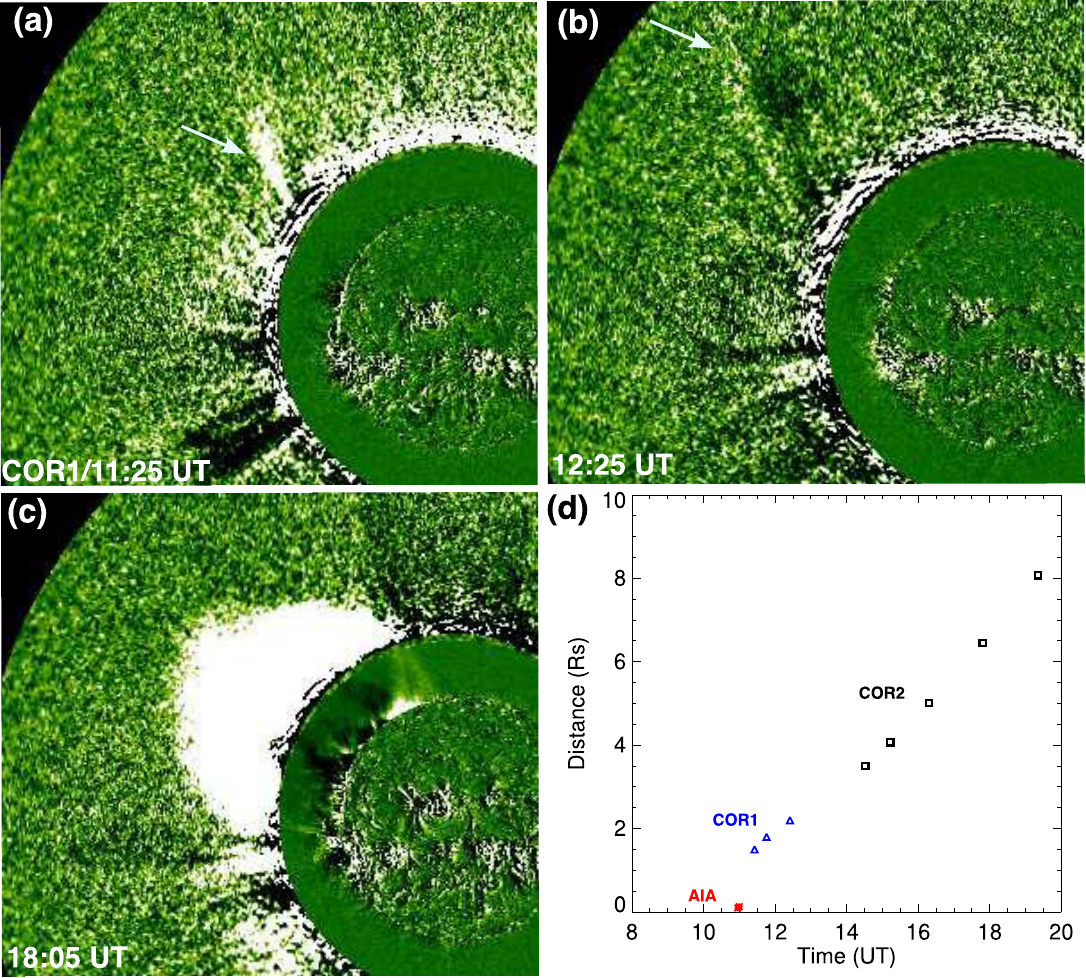}
}
\caption{(a–c) STEREO-A COR1 images showing the blob (marked by white arrows) and CME originating from the nested null-point topology. (d) The height-time evolution of the blob as observed in AIA, COR1, and COR2 images.} 
\label{app-fig3}
\end{figure*}

The active region was located behind the eastern limb in STEREO-A EUVI images. A dense blob was detected in STEREO-A COR1 images between 11:25 and 12:25~UT (Figure~\ref{app-fig3}(a,b)). The height-time plot of the blob tracing its motion from the low corona to the outer corona between 10:21 and 19:30~UT, as observed by AIA, COR1, and COR2.

\clearpage
\clearpage
\clearpage

\section{Supplementary Materials}
This section contains supplementary movies to support the results. All supplementary movies are available in the Zenodo repository at doi:\href{https://doi.org/10.5281/zenodo.16944513}{10.5281/zenodo.16944513}. \\
{\bf Movie S1}: An animation of the AIA 211~{\AA} intensity and running-difference images showing the X1.0 flare and associated coronal mass ejection. The right panel shows a larger field of view, while the left panel displays a zoomed-in view of the area outlined by the red box in the right panel. The animation runs from 17:03:01 UT to 18:16:25 UT. Its real-time duration is 14.9 s. \\
{\bf Movie S2}: An animation of the AIA 211~{\AA} intensity and running-difference images shows inflows and reconnection signatures near outer null N2. During the reconnection episode, downflows were observed along with the formation and ejection of the blob detected in STEREO-A COR1 and LASCO C2. The animation runs from 09:53:49 UT to 11:11:25 UT. Its real-time duration is 7.8 s. \\

\end{document}